\documentclass[
reprint, 
onecolumn,
prd, aps
]{revtex4-2}

\usepackage{graphicx}
\usepackage{dcolumn}
\usepackage{bm}
\usepackage{rotating}  
\usepackage{changepage}
\usepackage{amsmath}
\usepackage{tabularx}
\usepackage{amssymb}
\usepackage{booktabs}

\begin{document}

\title{Late-Time Cosmic Acceleration from QCD Confinement~Dynamics}
\thanks{Corresponde: humberto.martinezhuerta@udem.edu}%

\newcommand{\orcidauthorA}{0000-0003-0405-9344} 
\newcommand{\orcidauthorB}{0000-0002-6356-8870} 
\newcommand{\orcidauthorC}{0000-0001-7714-0704} 
\newcommand{\orcidauthorD}{0000-0002-0924-5843} 
\newcommand{\orcidauthorE}{0000-0001-7317-8211} 

\author{Jonathan Rinc\'on Saucedo}
\email{jonathan.rincon@udem.edu}
 
\author{H. Mart\'inez-Huerta*}%
\email{humberto.martinezhuerta@udem.edu}
\affiliation{Departamento de Física y Matemáticas, Universidad de Monterrey,\\Av. Morones Prieto 4500, 66238, San Pedro Garza Garc\'ia NL, México 
}%

\author{A. Huet}
\email{adolfo.huet@gmail.com}

\author{A. Hern\'andez -Almada}
\email{ahalmada@uaq.mx}
\affiliation{Facultad de Ingenier\'ia, Universidad Aut\'onoma de Quer\'etaro, Centro Universitario Cerro de las Campanas, Santiago de Queretaro 76010, Mexico}
\author{Miguel A. Garc\'ia-Aspeitia}
\email{angel.garcia@ibero.mx}
\affiliation{Departamento de F\'isica y Matem\'aticas, Universidad Iberoamericana Ciudad de M\'exico, Prolongaci\'on Paseo de la Reforma 880, Mexico City, 01219}

\begin{abstract}
We explore a phenomenological extension of the Polyakov–Nambu–Jona-Lasinio (PNJL) model by introducing a curvature-sensitive effective contribution to the Polyakov loop potential, motivated by the hypothesis that the non-perturbative QCD vacuum in the confined phase may retain a residual sensitivity to cosmic expansion. In a spatially flat FLRW background, this modification reduces to a term proportional to $\alpha(H/H_0)^d f(\Phi, \Phi^*)$, which naturally vanishes in the deconfined regime and behaves as an effective dynamical vacuum component at late times, without invoking a fundamental cosmological constant. The construction provides an effective thermodynamic description of the QCD sector within an adiabatic framework and introduces a minimal phenomenological extension characterized by the exponent $d$ and the amplitude parameter $\alpha$. We analyze the cosmological implications at the background level and confront the model with low-redshift observations, including cosmic chronometers, Type Ia supernovae, HII galaxies, and quasars. Using Bayesian Monte Carlo techniques, we constrain the model parameters and compare its performance with $\Lambda$CDM. Our results indicate that the modified PNJL cosmology provides a statistically competitive fit to current data while allowing small departures from $\Lambda$CDM within observational uncertainties. We also investigate the impact of the coupling on the QCD phase diagram and the critical end point. The framework offers a tractable effective approach to connect confinement physics with late-time cosmology and suggests directions for further theoretical development in QCD under curved backgrounds.
\end{abstract}

\keywords{cosmology; quark--gluon plasma; dark energy; finite-temperature field theory}
\maketitle


\section{Introduction}

Modern cosmology relies on General Relativity (GR) and the cosmological principle of homogeneity and isotropy to describe the large-scale structure and dynamics of the Universe. Within this framework, the Friedmann--Lemaître--Robertson--Walker (FLRW) metric and the associated Friedmann equations provide a robust foundation for modeling cosmic expansion. By incorporating various energy components—radiation, baryonic matter, neutrinos, dark matter (DM), and dark energy (DE)—the standard model, known as the $\Lambda$CDM, has proven to be remarkably successful in describing cosmological observations.

The inclusion of a cosmological constant $\Lambda$ in Einstein's field equations accounts for the observed late-time accelerated expansion of the Universe, initially discovered through type Ia supernova measurements~\cite{Riess:1998,Perlmutter:1999}. In the $\Lambda$CDM model, dark energy is characterized by an equation of state $w = -1$, corresponding to a vacuum energy that remains constant in time and space. However, this interpretation introduces profound theoretical challenges. The most notable is the so-called cosmological constant problem: Quantum field theory predicts vacuum energy densities that exceed the observed value by up to 120 orders of magnitude~\cite{Zeldovich, Weinberg}. Alternative interpretations of vacuum energy, based on effective models or thermodynamic considerations, have also been explored~\cite{Alcantara-Perez:2023jbv}.

To address these tensions, the cosmological community has proposed a wide range of dynamical dark energy models, including parameterized frameworks such as the phenomenologically emergent dark energy (PEDE) and generalized emergent dark energy (GEDE)~\cite{Li:2020ybr,Koo:2020ssl,Hernandez-Almada:2024ost,Hernandez-Almada:2020uyr}. Other approaches include models based on holographic and entropic principles~\cite{Barrow:2020tzx,Abreu:2021avp}, variable curvature scenarios~\cite{Esteban-Gutierrez:2024rpz}, and a range of theoretical proposals, including modifications to GR. These models aim to explain cosmic acceleration without invoking a fundamental cosmological constant. For comprehensive reviews covering dynamical, geometrical, and modified gravity approaches, see~\cite{Motta:2021hvl,DiValentino:2025sru}.

Recent observational analyses further motivate these efforts. For instance, results from the Dark Energy Spectroscopic Instrument (DESI) have hinted at deviations from a pure $\Lambda$CDM expansion, favoring scenarios with a mildly dynamical dark energy component~\cite{DESI:2025zgx}. This ongoing debate highlights the need for novel perspectives grounded in well-established physics.

In this context, the role of strong interactions becomes particularly intriguing. Quantum Chromodynamics (QCD), the gauge theory of the strong nuclear force, governs the behavior of strongly interacting matter and features a rich vacuum structure shaped by phenomena such as confinement and spontaneous chiral symmetry breaking. These features are most prominent in the nonperturbative regime, especially relevant in the early Universe, where the transition from a deconfined quark--gluon plasma to confined hadronic matter occurred.

The thermodynamics of this transition can be effectively modeled using the Polyakov--Nambu--Jona-Lasinio (PNJL) model~\cite{Mukherjee2010, Nishimura2010, Fukushima2004}, which extends the NJL framework by including the Polyakov loop as an order parameter for confinement. The PNJL model has been widely used to study the QCD phase diagram, including the location of the critical end point (CEP) and the interaction between chiral symmetry and color confinement~\cite{FukushimaSkokov2017, Gattringer2011}.

Despite its success in describing finite-temperature QCD, the PNJL model is usually treated in isolation from cosmological dynamics. However, some authors have speculated that the expansion of the Universe could influence the QCD vacuum structure~\cite{Holdom:2010ak, Urban:2009yg, Ohta:2010in}, potentially inducing effective contributions to dark energy. These ideas are typically inspired by non-perturbative effects or topological vacuum fluctuations in curved spacetimes.

Motivated by these considerations, we propose a phenomenological extension of the Polyakov--Nambu--Jona-Lasinio (PNJL) model by introducing a coupling between the Polyakov-loop potential and the Hubble parameter $H(t)$. The coupling term, proportional to $H^d$, will introduce a power-law sensitivity to the expansion rate of the Universe, where the exponent $d$ encapsulates the strength and nature of the cosmological backreaction on the QCD vacuum. This term is suppressed in the deconfined phase through a function $ f(\Phi,\Phi^*)$ that ensures the modification is active only in the confined regime.
This approach reflects the hypothesis that the strongly coupled QCD vacuum, dominant in the confined phase, could be sensitive to the large-scale expansion of the Universe, thus acting as an effective, dynamical component of dark energy.

We explore the cosmological implications of this modification by deriving the resulting Friedmann equations and constraining the model parameters using a combination of cosmic chronometers, type Ia supernovae, quasars, and HII galaxies. In parallel, we investigate the impact of the proposed coupling on the QCD phase diagram, including its effect on chiral and deconfinement transitions, and the location of the critical end point (CEP).

This paper is organized as follows. In Section~\ref{Sec: PNJL}, we review the standard PNJL model and its role in the description of QCD thermodynamics. In Section~\ref{Sec:modified PNJL}, we introduce the coupling to the Hubble parameter and justify its functional form. Section~\ref{sec:cosmology} derives the modified cosmological equations caused by the chromodynamic interactions. In Section~\ref{sec:constraints}, we describe the datasets used for parameter estimation, and Section~\ref{Results} presents our cosmological and QCD results. Finally, in Section~\ref{SD}, we summarize our findings and discuss future directions\footnote{We henceforth use units in which $c=\hbar=k_B=1$.}. 

\section{PNJL Model of QCD} \label{Sec: PNJL}

The Polyakov--Nambu--Jona-Lasinio model is a highly effective field theory employed to elucidate the non-perturbative regime of QCD. It extends the Nambu--Jona-Lasinio (NJL) model by incorporating the Polyakov loop, which serves as an order parameter for the confinement--deconfinement transition at finite temperatures. This model provides a unified framework to study both the chiral symmetry breaking and the confinement properties of QCD, thereby positioning it as a potent instrument for unraveling the thermodynamic behavior of strongly interacting matter \cite{Mukherjee2010, Nishimura2010}.

The Lagrangian density for the two-flavor PNJL model in the \( \mathrm{SU}(2) \) sector is given by~\cite{Ratti2006, Abuki2008}
\begin{equation}
\mathcal{L}_{\mathrm{PNJL}} =
\bar q \left(i\gamma^\mu D_\mu - m_0 + \gamma^0 \mu \right) q
+ \frac{G}{2} \left[(\bar q q)^2 + (\bar q i\gamma_5 \vec{\tau} q)^2 \right]
- \mathcal{U}(\Phi,\Phi^*,T),
\label{LagrangianPNJL}
\end{equation}
where \( D_{\mu} = \partial_{\mu} - i A_{\mu} \) is the covariant derivative, incorporating the coupling to the background gluonic field. In the context of the PNJL model in QCD, the covariant derivative includes a static temporal gluon background field \( A_0 \), introduced to model the coupling of quarks to the Polyakov loop. This background field is not dynamic; its function is to encode the thermal expectation value of the Wilson loop in imaginary time, which defines the traced Polyakov loop \( \Phi \). The spatial components of \( A_\mu \) are neglected, and only \( A_0 \) contributes to the temporal gauge.
The parameter \( m_{0} \) denotes the current quark mass, assumed equal for the up and down quarks due to isospin symmetry. 
The constant \( G \) is the effective coupling, and \( \mu \) is the quark chemical potential. 
The matrices \( \boldsymbol{\tau} \) are the Pauli matrices acting in flavor space, and \( \gamma_{5} \) is the usual Dirac matrix associated with chiral structure. 
Finally, \( \mathcal{U}(\Phi, T) \) represents the effective Polyakov-loop potential, which depends on the traced Polyakov loop \( \Phi \) and the temperature \( T \).

The Polyakov loop, \( \Phi \), is a crucial order parameter in effective QCD models, characterizing the breaking of the center symmetry of the \( \mathrm{SU}(3) \) color gauge group. In this work, we consider two light quark flavors (\( N_f = 2 \)) and three colors (\( N_c = 3 \)), following the standard setup of effective PNJL models. This approach enables the PNJL model to accurately characterize the transition from a confined hadronic state to a deconfined quark--gluon plasma. Effective QCD models at finite temperature and density are formulated through the thermodynamic potential as \cite{Hansen2007, Ghosh2006}.

\begin{equation}
\begin{split}
\Omega_{\mathrm{PNJL}}(T,\mu;M,\Phi,\Phi^*)
&= \mathcal{U}(\Phi,\Phi^*,T)
+ \frac{(M-m_0)^2}{4G}
- 2N_f \int \frac{d^3p}{(2\pi)^3}\, E_p
\nonumber\\
&
- 2N_f T \int \frac{d^3p}{(2\pi)^3}
\ln\!\left[1 + 3\Phi e^{-\beta(E_p-\mu)}
+ 3\Phi^* e^{-2\beta(E_p-\mu)}
+ e^{-3\beta(E_p-\mu)}\right]
\nonumber\\
&
- 2N_f T \int \frac{d^3p}{(2\pi)^3}
\ln\!\left[1 + 3\Phi^* e^{-\beta(E_p+\mu)}
+ 3\Phi e^{-2\beta(E_p+\mu)}
+ e^{-3\beta(E_p+\mu)}\right].
\label{termodinamic_potential_qcdL}
\end{split} 
\end{equation}
The term \( E_{\boldsymbol{p}} = \sqrt{\boldsymbol{p}^{2} + M^{2}} \) represents the quark energy, while \( M \) denotes the dynamically generated quark mass. In practice, the Polyakov-loop potential \( \mathcal{U}(\Phi,T) \) cannot be derived from first principles and is instead modeled through effective forms that capture the thermodynamic behavior of QCD near the confinement–deconfinement transition. Following Refs.~\cite{Ratti2006, Fukushima2008b}, we adopt a used parameterization fitted to reproduce lattice QCD results at zero and finite chemical potential. Thus, the form used in this work is given by,
\begin{equation}
\mathcal{U} = T^{4} \Bigg( -\dfrac{b_{2}(T)}{2} \Phi \Phi^{*} - \dfrac{b_{3}}{6} \big( \Phi^{3} + \Phi^{* 3} \big) + \dfrac{b_{4}}{4} \big( \Phi \Phi^{*} \big)^{2} \Bigg),
\end{equation}
where the temperature-dependent coefficients \( b_{2}, b_{3}, b_{4} \) control the behavior of the Polyakov-loop potential with parameters $a_{0} = 6.76$, $a_{1} = -1.95$, $a_{2} = 2.625$, $a_{3} = -7.44$, \mbox{$T_{0} = 270$ MeV},  $b_{3} = 0.75$, $b_{4} = 7.5$ and 
\begin{equation}
    b_{2} = a_{0} + a_{1} \Big(\dfrac{T_{0}}{T} \Big) +a_{2}\Big(\dfrac{T_{0}}{T} \Big)^{2} + a_{3} \Big( \dfrac{T_{0}}{T} \Big)^{3} \ .
\end{equation}
Unlike the NJL model, which lacks a direct description of confinement, the PNJL model provides a more complete thermodynamic picture by incorporating the effects of a background gluon field. The effective potential \( \mathcal{U} \) captures essential features of the deconfinement transition, while the NJL interaction term accounts for spontaneous chiral symmetry breaking. This unified treatment enables the PNJL model to describe the transition from hadronic matter to quark--gluon plasma more accurately. 

It is important to emphasize that the PNJL model should be understood as an effective framework that captures key qualitative features of QCD, rather than a fundamental description. In this work, we adopt a phenomenological perspective and explore the possible implications of coupling the QCD vacuum structure to the cosmological expansion rate. This approach is intended as an exploratory extension, aimed at identifying potential connections between confinement dynamics and cosmological evolution, rather than as a direct derivation from first-principles QCD.

In this context, the PNJL framework provides a tractable setting in which such a hypothesis can be implemented and studied. In the next section, we introduce a phenomenological extension that incorporates this coupling through the Hubble parameter.

\section{Modified PNJL Model} \label{Sec:modified PNJL}

The Polyakov loop, \( \Phi \), plays a central role in this framework, serving as a fundamental order parameter in effective QCD models. It characterizes the spontaneous break of the \( \mathbb{Z}(3) \) center symmetry of the \( \mathrm{SU}(3) \) color gauge group, which is closely related to the confinement–deconfinement transition~\cite{FukushimaSkokov2017, Gattringer2011}, and its dynamics are described by the effective Polyakov-loop potential $\mathcal{U}(\Phi,\Phi^{*},T)$, which summarizes the thermodynamic behavior of the strongly interacting medium near the transition region~\cite{Fukushima2004}.

Motivated by studies of the QCD vacuum in curved backgrounds and expanding
spacetimes~\cite{Holdom:2010ak, UrbanZhitnitsky2011, Ohta:2010in}, we explore the possibility that confinement dynamics encoded in $\mathcal{U}(\Phi,\Phi^{*},T)$ may be weakly sensitive to the large-scale expansion of the Universe. These analyses suggest that vacuum
condensates and topological fluctuations of strongly interacting fields can receive
curvature- or expansion-dependent corrections in FLRW geometries. Although these effects
are typically very small at present times, they may nonetheless induce an effective
contribution to the dark sector.

Guided by these ideas, we introduce a minimal, phenomenological coupling between the
Polyakov-loop potential and the Hubble parameter $H(t)$. Both the QCD
scale $\Lambda_{\rm QCD}$ and $H(t)$ have dimensions of mass, in natural units, which allows us to construct a dimensionally consistent modification of the effective potential without resorting to
higher-order curvature invariants. To keep track of the overall scale, we introduce a
constant $\alpha$ with dimensions of energy density (${\rm MeV}^4$), while the dimensionless
ratio $H(t)/H_{0}$ controls the relative strength of the correction at a given cosmic
epoch, with $H_{0}$ the present-day Hubble~\mbox{parameter}.

The modified Polyakov-loop potential is then expressed as
\begin{equation}
\mathcal{U}'(\Phi,\Phi^{*},T,H(t))
 \;=\;
\mathcal{U}(\Phi,\Phi^{*},T)
 \;+\;
\alpha \left(\frac{H(t)}{H_{0}}\right)^{d} f(\Phi,\Phi^{*}) ,
\label{eq:Uprime}
\end{equation}
where $\mathcal{U}(\Phi,\Phi^{*},T)$ is the standard PNJL potential,
$\alpha$ controls the amplitude of the cosmological correction and is fixed to $\alpha = 200\,\mathrm{MeV}^4$ throughout the analysis, $d$ is a real
exponent obtained from the cosmological fits that parametrizes how the QCD vacuum responds to the expansion rate, and
$f(\Phi,\Phi^{*})$ is a dimensionless function that modulates the effect of the coupling
in different phases of QCD.

The new term in Equation~\eqref{eq:Uprime} is designed to satisfy two basic requirements: (a) it should be relevant only in the confined phase, where non-perturbative vacuum effects dominate; and (b) it should be strongly suppressed in the deconfined phase, in order not to interfere with the thermodynamics of the quark--gluon plasma and with early-Universe physics such as Big Bang nucleosynthesis and the CMB. These conditions are implemented through the function $f(\Phi,\Phi^{*})$, which we choose to depend only on the gauge-invariant combination $\Phi\Phi^{*}$. In particular, we adopt the simple quadratic form
\begin{equation}
f(\Phi,\Phi^{*}) = \left(1 - \Phi\Phi^{*}\right)^{2} ,
\label{eq:fPhi}
\end{equation}
which satisfies
\begin{equation}\label{eq:confinedphase}
f(\Phi,\Phi^{*}) \to 1 \quad \text{for} \quad |\Phi| \to 0 
\qquad \text{(confined phase)} ,
\end{equation}
and
\begin{equation}
f(\Phi,\Phi^{*}) \to 0 \quad \text{for} \quad |\Phi| \to 1 
\qquad \text{(deconfined phase)} .
\end{equation}
This suppression mechanism is not implemented explicitly in the cosmological background equations used in this work, but rather provides the physical motivation for restricting the effective description to the low-redshift regime where $f(\Phi,\Phi^{*}) \simeq 1$.

The derivatives of $f$ with respect to $\Phi$ and $\Phi^{*}$,
\begin{equation}
\frac{\partial f}{\partial \Phi} = -2 \Phi^{*}\left(1 - \Phi\Phi^{*}\right), 
\qquad
\frac{\partial f}{\partial \Phi^{*}} = -2 \Phi \left(1 - \Phi\Phi^{*}\right),
\end{equation}
are regular and do not introduce singularities in the equations of motion for the
mean fields. Therefore, the modified potential $\mathcal{U}'$ remains smooth across the
transition region.

With this construction, the influence of the Hubble parameter is maximal when the system
is deeply confined ($\Phi \simeq 0$), where the QCD vacuum structure is dominated by
non-perturbative condensates, and it gradually switches off as the system approaches the
deconfined phase ($\Phi \simeq 1$). This matches the physical expectation that any
cosmological backreaction on the QCD vacuum should be most relevant in the strongly
coupled regime, while becoming negligible in the quark--gluon plasma.

The modified term in Equation~\eqref{eq:Uprime} also induces a corresponding change in the
total thermodynamic potential,
\begin{equation}
\Omega_{\rm PNJL}' \;=\; \Omega_{\rm cond} + \Omega_{\rm quarks}
+ \mathcal{U}'(\Phi,\Phi^{*},T,H(t)) ,
\end{equation}
which in turn affects the chiral and deconfinement transitions and the location of the
critical end point in the phase diagram. In addition, when evaluated at a fixed cosmic
epoch, the extra contribution proportional to $H^{d}$ can be interpreted as an effective,
dynamical dark-energy component in the Friedmann equations, as discussed in the next
section.

\subsection*{Adiabatic Approximation and Cosmological Consistency}\label{sec:Adiabatic}

In our framework, the QCD and cosmological sectors are coupled 
through the modified Polyakov-loop potential introduced in Equation~\eqref{eq:Uprime}. However, the respective energy scales differ by many orders of magnitude. The characteristic scale governing QCD
thermodynamics is $\Lambda_{\rm QCD} \sim 200\,\mathrm{MeV}$, whereas the present Hubble
parameter is extremely small, $H_{0}\sim10^{-33}\,\mathrm{eV}$. The ratio,
\begin{equation}
\frac{\Lambda_{\rm QCD}}{H_{0}} \sim 10^{41},
\end{equation}
implies that the cosmic expansion is effectively static from the point of view of QCD
dynamics. This enormous hierarchy allows an adiabatic approximation, in which
$H(t)$ acts as an external parameter: the thermodynamic potential of the PNJL model is
evaluated at a fixed cosmological epoch, without requiring a dynamical backreaction on the
gap~equations.

This separation of scales justifies treating $H(t)$ as a slow, external variable in the QCD sector. Within this adiabatic approximation, the thermodynamic potential is evaluated at a fixed but representative value of the expansion rate, rather than being dynamically evolved along the full cosmological history. In practice, the modification is implemented through the factor $(H/H_0)^d$, which is treated as an effective parameter encoding the influence of the cosmological background. This allows us to explore how different values of $d$ modify the thermodynamic behavior of the model within a controlled and computationally tractable~setup.

At the same time, although the PNJL thermodynamics are computed at fixed $H(t)$, the parameter $d$ appearing in the modification is directly inherited by the cosmological sector, entering the Friedmann equation as a contribution proportional to $H^{d}$. The cosmological implications of this term—including its interpretation as a dynamical dark-energy component and its effect on the Hubble expansion history—are analyzed in detail in the next~section.

This interpretation is consistent with previous studies in which $H(t)$ is introduced as an
external parameter encoding the influence of cosmic expansion on QCD vacuum
contributions~\cite{Holdom:2010ak, UrbanZhitnitsky2011}. Within this adiabatic scheme, the QCD sector can be treated as effectively thermodynamic, while the cosmological evolution retains a phenomenological imprint of the strongly coupled regime through the exponent $d$.

\section{The Cosmology} \label{sec:cosmology}

The modification introduced in Equation~\eqref{eq:Uprime} adds to the QCD thermodynamic potential a contribution proportional to $H^{d}(t)$, modulated by the function $f(\Phi)$. When evaluated at a fixed cosmic epoch, this term behaves as a vacuum-like energy density.
In this section, we show explicitly how this contribution enters the Friedmann equations and
how the parameter $d$ characterizes the resulting cosmological dynamics.

\subsection{From the Modified PNJL Potential to the Friedmann Equation}

At a given epoch $t$, the additional contribution to the effective potential,
\begin{equation}
\Delta \mathcal{U}(t) = \alpha \left(\frac{H(t)}{H_0}\right)^{d} f(\Phi,\Phi^{*}),
\label{eq:deltaU}
\end{equation}
acts as a homogeneous energy density in the stress--energy tensor. In the confined phase,
where $\Phi \simeq 0$, and because \eqref{eq:confinedphase}, we have $f(\Phi,\Phi^{*}) \to f(0)=1$. Thus, it is possible to define
\begin{equation}
    C\equiv\frac{\alpha}{H_0^d}.
\end{equation}
Therefore, we associate $\Delta\mathcal{U}(t)$ with an effective energy density, $\rho_{\text{QCD}}(t)=C H(t)^d$, and introduce a corresponding energy--momentum tensor $T_{\mu\nu}^{\text{QCD}}$, treating it as a perfect fluid at the background level.
Consistency with the Einstein equations, together with the Bianchi identities, implies the conservation condition $\nabla^{\mu}T_{\mu\nu}^{\text{QCD}}=0$ within this effective description. This requirement leads to $p_{QCD}=-\frac{C}{3}\left[dH^{d-2}\dot{H}+3H^d\right]$, with equation of state 
\begin{equation}
    w_{QCD}\equiv\frac{p_{QCD}}{\rho_{QCD}}=-\frac{dH^{d-2}\dot{H}+3H^d}{3H(t)^d}.
\end{equation}
In this case, our Einstein field equation takes the form
\begin{equation}
    G_{\mu\nu}=8\pi G[T_{\mu\nu}^{stand}+T_{\mu\nu}^{QCD}], \label{eq:Einstein}
\end{equation}
where stand and QCD refer to the stress--energy tensor of standard fluids (baryons, radiation, etc.) and QCD contributions, respectively, associated with perfect fluids, and $G_{\mu\nu}$ is the Einstein tensor.

Then, both contributions of the stress--energy tensor follow the standard continuity equation, with the following expression
\begin{equation}
    \sum_i\left[\dot{\rho}_i+3H(\rho_i+p_i)\right]=0.
\end{equation}
The continuity equation is valid for  standard fluids and the QCD contribution that transits from the PNJL EoS at QCD scales (see details in Section \ref{Sec:modified PNJL}).

Solving Equation \eqref{eq:Einstein}, the Friedmann equation yields
\begin{equation}
    H^2 (t) \equiv\left(\frac{\dot{a}}{a}\right)^2=\frac{8\pi G}{3}\left[\sum_i\rho_i + \rho_{\rm QCD}(t) \right] = \frac{8\pi G}{3}\left[\sum_i\rho_i +C\, H^{d}(t)\right].
\label{eq:FriedmannHd}
\end{equation}
The new term acts as an effective, dynamical dark-energy density whose time evolution is determined by the expansion rate itself. The standard $\Lambda$CDM model is recovered in the limit $d=0$,
for which the correction reduces to a constant value $C$, which originates from the modified PNJL potential, and for $d\neq0$ we have a term that evolves like a dynamical dark energy.

This approach preserves the standard form of General Relativity by incorporating the QCD-induced term as an additional energy component in the Friedmann equations, while keeping the continuity equation unaltered. As can be observed, the parameter $d$ is inherited from Equation \eqref{eq:Uprime}, uncovering information on QCD eras.

\subsection{The Dimensionless Friedmann Equation}

Introducing the normalized Hubble parameter $E(z)=H(z)/H_{0}$ and the usual density
parameters $\Omega_{0m}$ and $\Omega_{0r}$, Equation~\eqref{eq:FriedmannHd} becomes
\begin{equation}\label{FriedmannModified}
    E^2(z)-\zeta E(z)^d=\Omega_{0m}(z+1)^3+\Omega_{0r}(z+1)^4, 
\end{equation}
where the dimensionless parameter
\begin{equation}\label{eq:zeta=c}
\zeta \equiv \frac{8\pi G}{3} CH_0^{d-2}=\frac{8\pi G}{3} \left(\frac{\alpha }{H_0^{2}}\right),
\end{equation}
characterizes the strength of the QCD-induced contribution relative to $H_{0}^{2}$.

Evaluating Equation~\eqref{FriedmannModified} at $z=0$ gives the constraint
\begin{equation}\label{eq:1-zeta}
1 - \zeta = \Omega_{0m} + \Omega_{0r},
\end{equation}
which is the analogue of the standard Friedmann relation. For $d=0$, the term $\zeta$
plays the role of an effective cosmological constant.

\subsection{The Cosmographic Parameters}

To assess the impact of the parameter $d$ on the expansion, and to enable a
model-independent comparison with the $\Lambda$CDM, we compute three cosmographic quantities:
the deceleration parameter $q(z)$, the jerk parameter $j(z)$, and the effective equation
of state $w_{\rm eff}(z)$. These quantities depend only on derivatives of $H(z)$ and are
independent of the theoretical origin of the energy components.

The deceleration parameter is defined as
\begin{equation}\label{eq:q(z)}
q(z)=\frac{(z+1)}{2E(z)^2}\frac{dE(z)^2}{dz}-1,
\end{equation}
while the jerk parameter reads
\begin{equation}\label{eq:j(z)}
j(z) = q(z)\left[2q(z)+1\right] + (1+z)\frac{dq(z)}{dz}.
\end{equation}
Finally, the effective equation of state associated with the {total perfect cosmic fluid is
\begin{equation}\label{eq:w(z)}
w_{\rm eff}(z) \equiv \frac{p(z)_{\rm eff}}{\rho(z)_{\rm eff}}=\frac{1}{3}\left[2q(z)-1\right],
\end{equation}
where the subscript ``eff'' means that all  fluids are involved in the EoS.} Even when the observational constraints favor those values of $d$ close to zero, these
diagnostics remain sensitive to small deviations from the $\Lambda$CDM and therefore provide
a useful tool for identifying the dynamical character of the QCD-induced contribution
in Equation~\eqref{eq:FriedmannHd}.

The next section describes the datasets and statistical methods used to constrain
the parameter space $\{h,\,\Omega_{0m},\,d\}$ and to assess the viability of
the model compared to the standard $\Lambda$CDM cosmology.

\section{Datasets} \label{sec:constraints}

To constrain the free parameters of the QCD-modified cosmological model, we define the parameter space as $\Theta = \{ h,\; \Omega_{0m},\; d \}$, where $h$ is the dimensionless Hubble parameter, $\Omega_{0m}$ denotes the
total matter density parameters, and $d$ is the exponent
characterizing the dynamical behavior of the QCD-induced contribution $\zeta E^d(z)$ in
Equation~\eqref{FriedmannModified}. The parameter $\zeta$ is not an independent quantity,
as it is fixed by the Friedmann constraint at $z=0$ through Equations~\eqref{eq:zeta=c} and~\eqref{eq:1-zeta},
 with $\Omega_{0r}$ computed from the standard cosmic microwave background temperature.

Constraints on these parameters are obtained through a combination of recent observational datasets, including cosmic chronometers, Type Ia supernovae, hydrogen-II galaxies, and intermediate-luminosity quasars.  We perform statistical analysis using a Markov Chain Monte Carlo (MCMC) approach via the \texttt{Emcee} Python 3 package \cite{Foreman:2013}. To ensure chain convergence, we monitor the autocorrelation function and adopt 2000 chains of 200 steps each, adopting Gaussian priors for $h=0.6766 \pm 0.0042$ and $\Omega_{0m}=0.3111\pm 0.0056$, and a uniform prior for $-10<d<10$. We adopt these Gaussian priors on $h$ and $\Omega_{0m}$ with the aim of deriving more stringent constraints on the characteristic parameter $d$ of the QCD model by propagating their uncertainties into the inferred values of $d$, rather than fixing these parameters a priori. In the following, we summarize the dataset.

\begin{itemize}
    \item \textit{Cosmic chronometers} (CCs): A sample of 33 measurements of the Hubble parameter that covers a redshift region of $0.07<z<1.965$. The CC sample contains 15 correlated measurements and 18 points of $H(z)$ considered uncorrelated \cite{M_Moresco_2012, Moresco_2015, Moresco_2016, Moresco_2020, Jiao_2023, Tomasetti_2023}. {To constrain the space parameter of the model, the $\chi^2$-function is defined as
    \begin{equation}
        \chi^2_{\rm CC} = \sum_i^{18}\left( \frac{H_{obs}^i-H_{th}(z_i, \Theta)}{\sigma^i}\right)^2 + \Delta\vec{H}\,{\rm Cov}^{-1}\Delta\vec{H}^{T},
    \end{equation}
    where the sum in the first term runs over all uncorrelated points, while the second term operates over the correlated sample. Cov$^{-1}$ is the inverse of the covariance matrix of the vector $\Delta\vec{H}= \vec{H}_{obs}-\vec{H}_{th}$, corresponding to the difference between the observed and theoretical vectors of $H$.}
    \item \textit{Type Ia supernovae} (SNIa): The Pantheon+ dataset \cite{Scolnic2018-qf, Brout_2022} contains 1701 correlated measurements of the distance modulus in the redshift region $0.001<z<2.26$. We use a function $\chi^2$ for correlated data to eliminate contributions {of nuisance parameters defined as \cite{Conley2010}
    \begin{equation}
        \chi^2_{\rm SNIa} = a + \log \left( \frac{e}{2\pi} \right) - \frac{b^2}{e},
    \end{equation}
    where 
    \begin{eqnarray}
        a &=& \Delta\boldsymbol{\tilde{\mu}}^{T}\cdot\mathbf{Cov_{P}^{-1}}\cdot\Delta\boldsymbol{\tilde{\mu}}, \nonumber\\
        b &=& \Delta\boldsymbol{\tilde{\mu}}^{T}\cdot\mathbf{Cov_{P}^{-1}}\cdot\Delta\mathbf{1}, \\
        e &=& \Delta\mathbf{1}^{T}\cdot\mathbf{Cov_{P}^{-1}}\cdot\Delta\mathbf{1}\, , \nonumber
    \end{eqnarray}
    and $\Delta\boldsymbol{\tilde{\mu}}$ is the vector of the difference between the theoretical and observed distance modulus, defined as
        \begin{equation}
    \mu_{th}(z, \Theta) = 5 \log_{10} \left [ \frac{d_L(z, \Theta)}{1\,{\rm Mpc}}\right] + 25,
    \end{equation}
    which is related to the luminosity distance $d_L$,  
        \begin{equation}\label{eq:dL}
    d_L(z)=(1+z)c\int_0^z\frac{dz^{\prime}}{H(z^{\prime})}\,,
    \end{equation}
    
    Here, $c$ is the speed of light, $\Delta\mathbf{1}=(1,1,\dots,1)^T$ is the transpose of the unit vector and $\mathbf{Cov_{P}}$ is the covariance matrix.}
    \item \textit{ydrogen II galaxies} (HIIGs): This sample includes 181 distance modulus measurements of low-mass ($M < 10^9 M_{\odot}$) compact systems with star-forming regions, covering $0.01 < z < 2.6$ \cite{GonzalezMoran2019, Gonzalez-Moran:2021drc}. To compare the QCD model to data, the $\chi^2$-function is built as
    \begin{equation}\label{eq:chi2_HIIG}
       \chi^2_{{\rm HIIG}}=\sum_i^{181}\frac{[\mu_{th}(z_i, \Theta)-\mu_{obs}^i]^2}{\epsilon_i^2}\,,
    \end{equation}
    where $\mu_{obs}^i \pm \epsilon_i$ is the distance modulus and its uncertainty observed at the redshift $z_i$ and $\mu_{th}$ is its theoretical quantity.
    \item \textit{Intermediate-luminosity quasars} (QSOs): Composed of 120 angular size measurements from ultra-compact radio sources in the region $0.462 < z < 2.73$ \cite{ShuoQSO:2017}, this dataset is analyzed with an uncorrelated function \(\chi^2\), marginalizing potential nuisance parameters related to the distance modulus. {Thus, an uncorrelated $\chi^2$-function is used as
    \begin{equation}
    \chi^2_{\rm QSO} = \sum_i^{120} \left( \frac{\theta_{obs}^i-\theta_{th}(z_i,\Theta)}{\sigma_{\theta_{obs}^i}}\right)^2\,,
    \end{equation}
    where $\theta_{obs}^i \pm \sigma_{\theta^i_{obs}}$ is the observed measurement  and its corresponding uncertainty at redshift $z_i$, and $\theta_{th}(z_i)$ is the theoretical counterpart estimated through the relation $\theta(z)=l_m/D_A(z)$, where $D_A(z)$ is the angular diameter distance related to the luminosity distance through $D_A(z)=d_L(z)/(1+z)^2$ and $l_m$ is an intrinsic length of the QSO fixed to $l_m = 11.03\pm 0.25\,$pc \cite{ShuoQSO:2017}.}
\end{itemize}

\section{Results} \label{Results}

\subsection{Cosmology Results}

The cosmological implications of the QCD-modified Friedmann equation are now examined using the datasets described in the previous section and covering a redshift region up to $z<2.73$ when the QSO sample is added. The parameters $\{h,\,\Omega_{0m},\,d\}$ are jointly constrained, while the constant
$\zeta$ is fixed by the $z=0$ Friedmann constraint.
Table~\ref{tab:bf_model} summarizes the resulting median values and 68\% ($1\sigma$) confidence intervals for several data combinations. In all cases, the best-fit values of the exponent $d$ cluster around zero, indicating that the QCD-induced term behaves very similarly to a cosmological constant. This is consistent with the expectation that the coupling to the Hubble rate should be small at the present epoch. Furthermore, the standard deviations of the posterior distributions for $h$ and $\Omega_{0m}$ remain essentially unchanged relative to their prior values, indicating that the available data provide only weak additional constraints on these parameters\footnote{{The covariance matrix of the posterior distribution can be written as $\Sigma_p = (\Sigma_0^{-1} + \Sigma_D^{-1})^{-1}$, where $\Sigma_0$ and $\Sigma_D$ denote the prior and likelihood covariance matrices, respectively. When the likelihood is weakly informative $\Sigma_D^{-1} \ll \Sigma_0^{-1}$, one naturally obtains $\Sigma_p \approx \Sigma_0$.}}. 
However, the constraints do not force $d$ to be exactly zero, and small deviations from $\Lambda$CDM remain allowed within 1$\sigma$.

To better understand the dynamical character of the model, we compute the cosmographic quantities $q(z)$, $j(z)$ and $w_{\rm eff}(z)$. These functions depend only on $H(z)$ and its derivatives and are therefore sensitive even to small variations in $d$. Importantly, they provide a model-independent diagnostic: their interpretation does not rely on the microscopic origin of the energy component but only on its impact on the background expansion. This justifies their use even when $d$ is close to zero, addressing potential
concerns about the relevance of Equations~\eqref{eq:q(z)} to \eqref{eq:w(z)}.

\begin{table}[h]
\small
	\caption{Median values and their $1\sigma$ confidence interval for the QCD cosmology and $\Lambda$CDM using CC, SNIa, HIIG and QSO datasets. {The standard deviations of the posterior distribution for $h$ and $\Omega_{0m}$ are dominated by their prior, as the likelihood provides only weak constraints.}}\label{tab:bf_model}
\begin{tabularx}{\textwidth}{cccccccc}
\toprule
\textbf{Data} & \boldmath{$\chi^2$}  & \boldmath{$h$} & \boldmath{$\Omega_{0m}$} & \boldmath{$d$} & \boldmath{$\tau_U \,[\rm{Gyrs}]$}  & \boldmath{$z_T $}  &  \boldmath{$q_0 $}  \\ 
\midrule
CC &16.52 & $0.678^{+0.004}_{-0.004}$  & $0.312^{+0.006}_{-0.006}$  & $-0.541^{+0.429}_{-1.179}$ & $14.021^{+0.318}_{-0.204}$ & $0.663^{+0.018}_{-0.022}$ & $-0.604^{+0.055}_{-0.102}$  \\ 
SNIa &2011.65 & $0.677^{+0.004}_{-0.004}$   & $0.317^{+0.005}_{-0.005}$  & $-0.005^{+0.028}_{-0.059}$ & $13.747^{+0.106}_{-0.104}$ & $0.628^{+0.013}_{-0.013}$ & $-0.527^{+0.009}_{-0.010}$ \\ 
CC+SNIa 
&2026.57 & $0.677^{+0.004}_{-0.004}$   & $0.318^{+0.005}_{-0.005}$  & $-0.004^{+0.027}_{-0.057}$ & $13.736^{+0.107}_{-0.106}$ & $0.627^{+0.013}_{-0.013}$ & $-0.526^{+0.009}_{-0.010}$ \\ 
CC+SNIa+HIIG &2467.77 & $0.679^{+0.004}_{-0.004}$  & $0.318^{+0.005}_{-0.005}$  & $-0.003^{+0.026}_{-0.057}$ & $13.697^{+0.098}_{-0.092}$ & $0.627^{+0.013}_{-0.013}$ & $-0.526^{+0.009}_{-0.011}$ \\ 
CC+SNIa+QSO &5197.86 & $0.684^{+0.004}_{-0.004}$  & $0.316^{+0.005}_{-0.005}$  & $-0.008^{+0.030}_{-0.063}$ & $13.618^{+0.108}_{-0.108}$ & $0.631^{+0.013}_{-0.014}$ & $-0.529^{+0.010}_{-0.012}$ \\
CC+SNIa+HIIG+QSO &5637.75 & $0.685^{+0.004}_{-0.004}$  & $0.316^{+0.005}_{-0.005}$  & $-0.010^{+0.030}_{-0.063}$ & $13.601^{+0.097}_{-0.099}$ & $0.633^{+0.013}_{-0.013}$ & $-0.530^{+0.010}_{-0.011}$ \\
\bottomrule
\end{tabularx}
\end{table}

\begin{figure}[h]
    \includegraphics[width=0.8\textwidth]{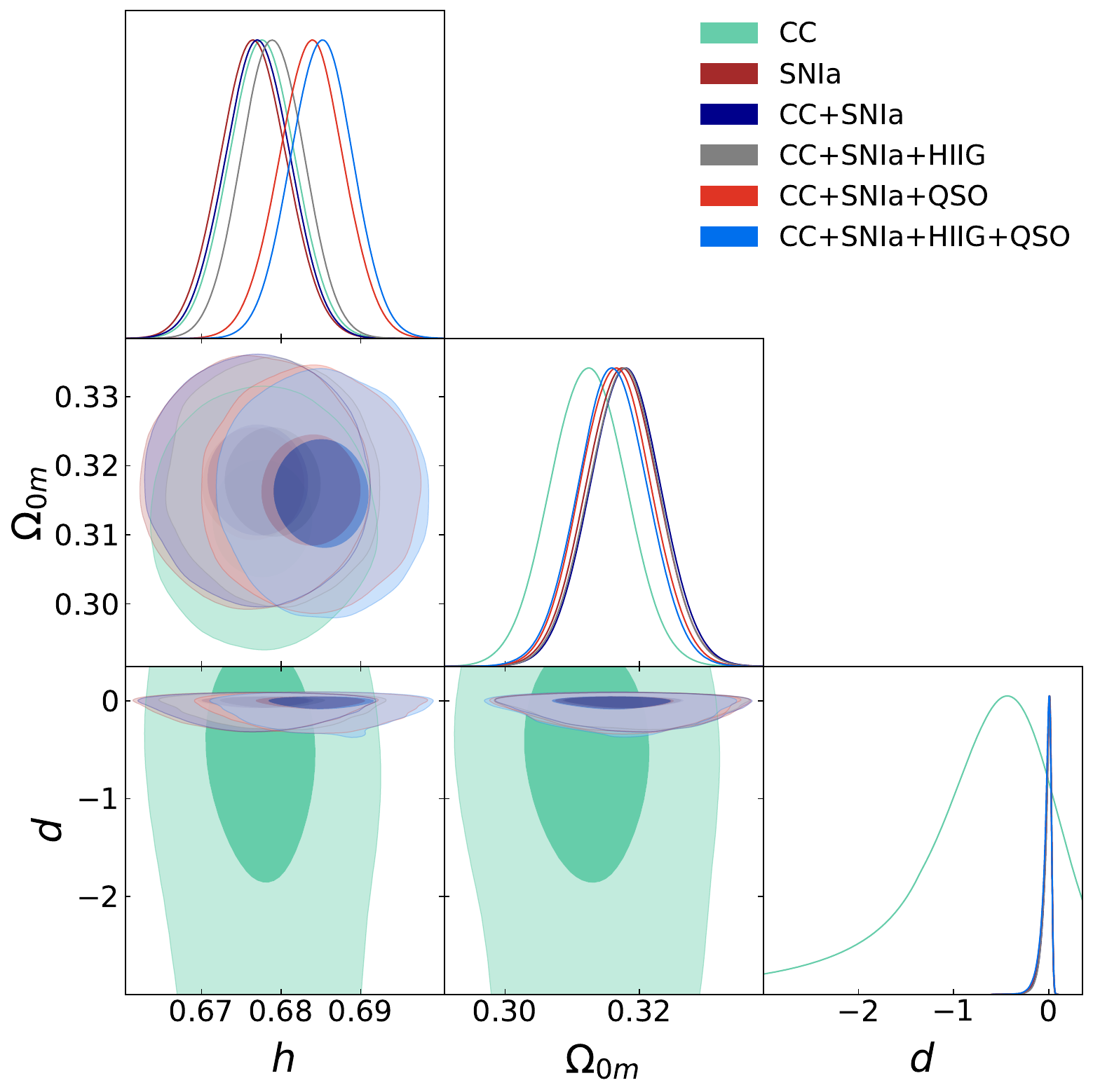}
    \caption{1D posterior distributions and 2D contours at $1\sigma$ (inner region) and $3\sigma$ (outermost region) CL for different data combinations for QCD model.}
    \label{fig:contours}
\end{figure}

Figure~\ref{fig:contours} shows the 1D posterior distributions and the 2D confidence contours at the confidence level $1\sigma$ and $99.7\%$ ($3\sigma$) for the parameters of the QCD model using multiple datasets. As we expected, the combination of CC and SNIa yields significantly tighter constraints, particularly in the parameter $d$ that quantifies deviations from behavior similar to $\Lambda$CDM. We do not observe strong degeneracies between $d$ and other cosmological parameters, with all datasets consistently favoring values clustered around $d = 0$.

\begin{figure}[h]
    \includegraphics[width=0.45\textwidth]{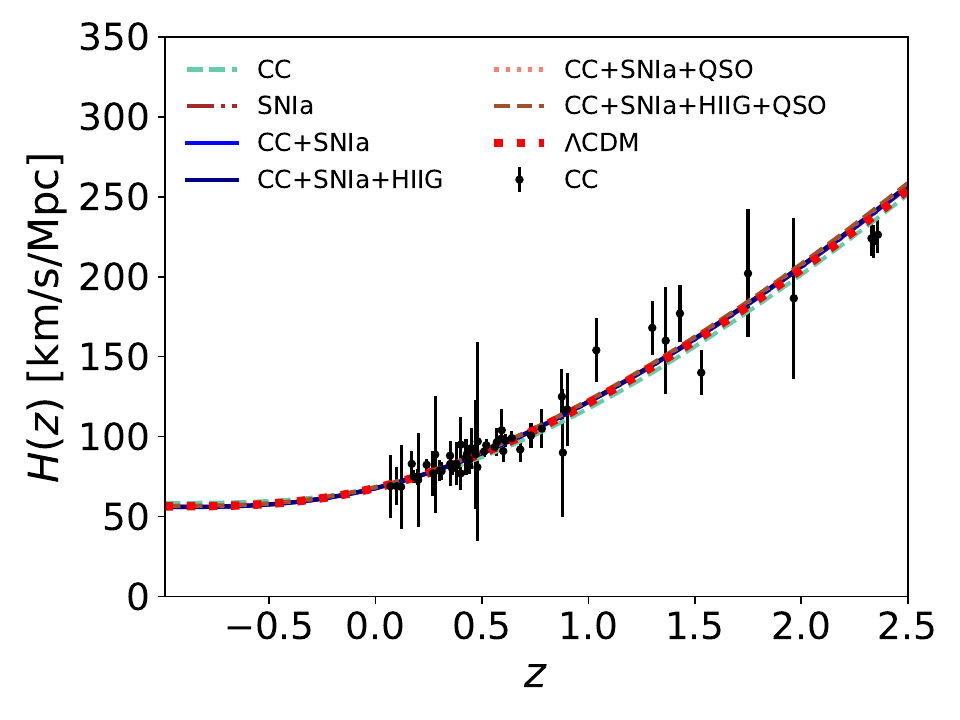}
    \includegraphics[width=0.45\textwidth]{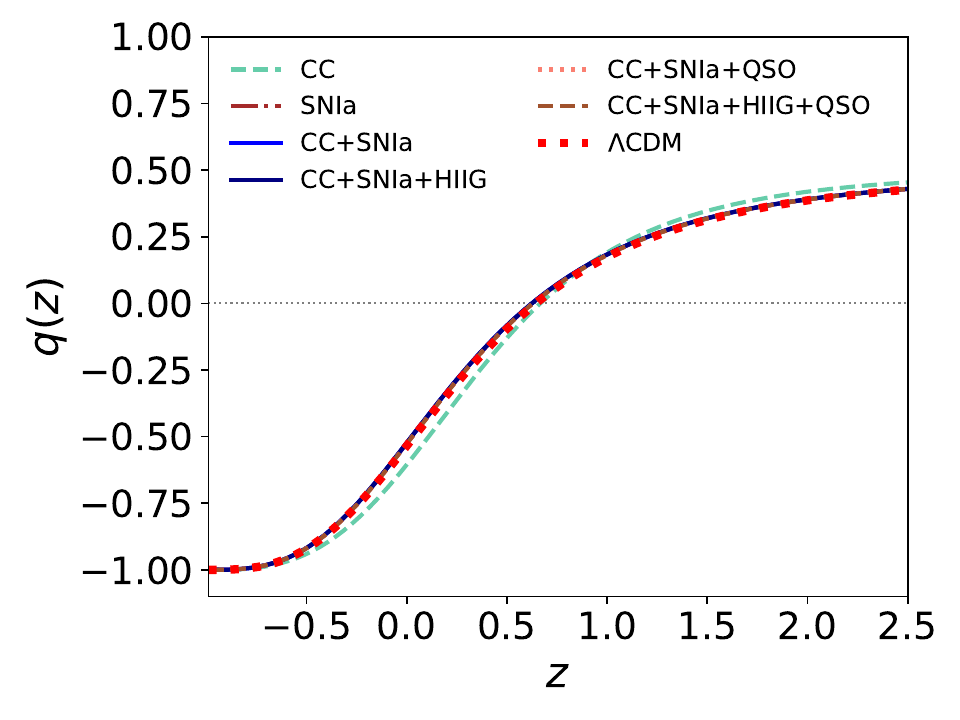}\\
    \includegraphics[width=0.45\textwidth]{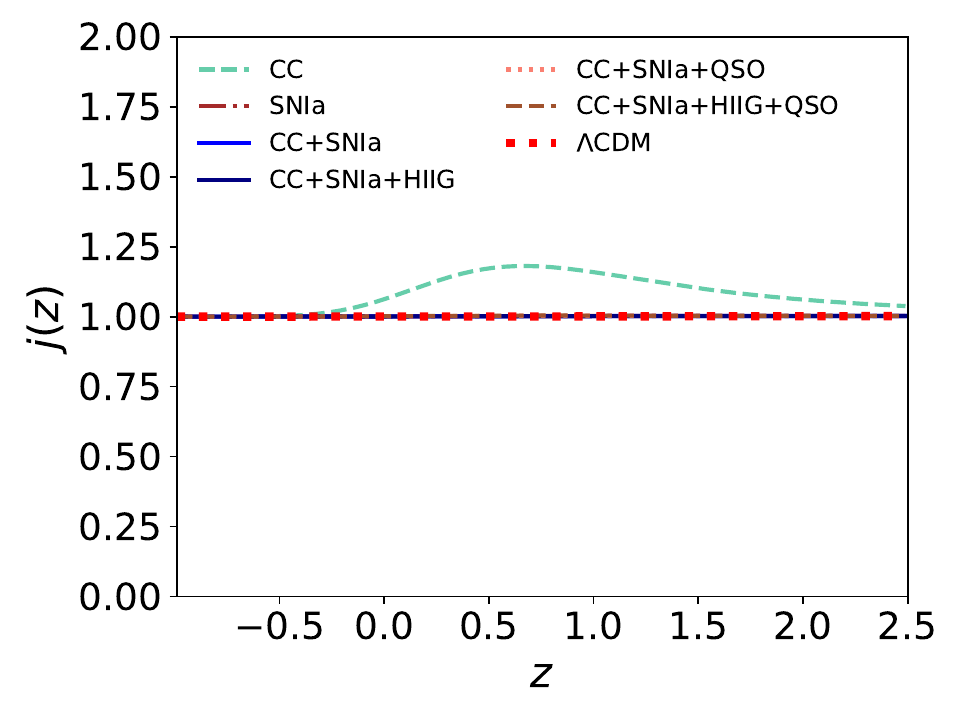}
    \includegraphics[width=0.45\textwidth]{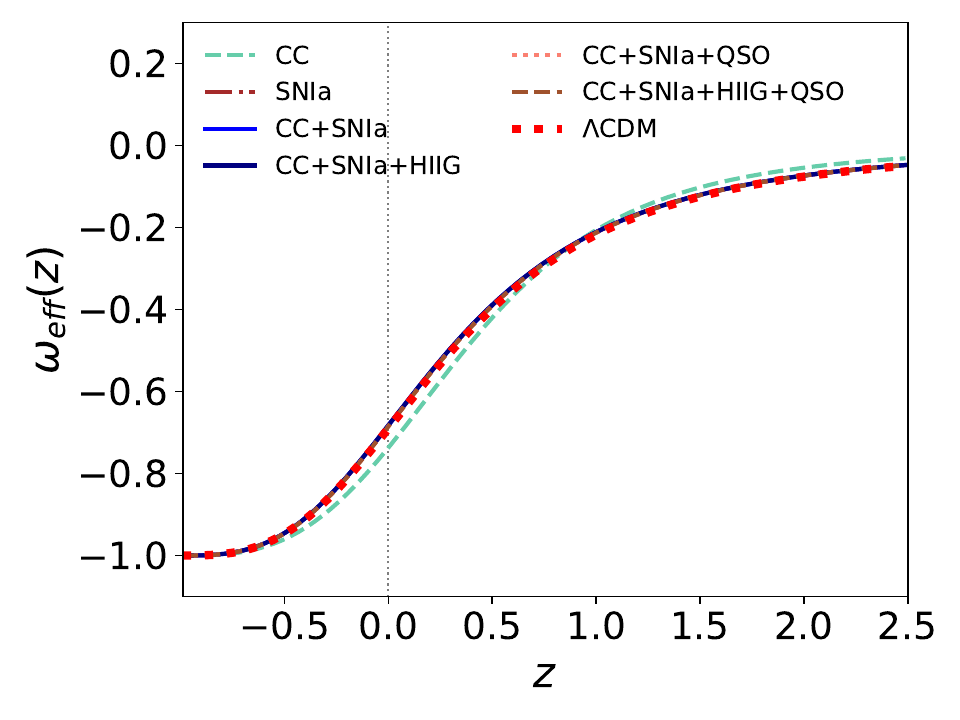}
    \caption{Reconstruction of the Hubble parameter, and the deceleration parameter (\textbf{top panel}), and jerk parameter and the effetive equation of state \linebreak (\textbf{bottom panel}) for different data combinations for the QCD model in the redshift range $-1<z<2.5$ using different data combinations. The standard $\Lambda$CDM  is included as red dashed lines.}
    \label{fig:cosmography}
\end{figure}

Figure~\ref{fig:cosmography} presents the reconstructed evolution of the Hubble parameter $H(z)$, the deceleration parameter $q(z)$, the jerk parameter $j(z)$, and the effective EoS for the QCD-modified cosmology, using several combined datasets. We find that the behavior aligns with the expectations of the $\Lambda$CDM at low redshifts, allowing only minor deviations. The deceleration parameter exhibits a smooth transition from deceleration to acceleration, while the jerk parameter remains close to the canonical value $j \simeq 1 $ as expected for a nearly constant dark-energy component and supporting the model's consistency with the observed expansion history.

\begin{figure}[h]
    \includegraphics[width=0.48\textwidth]{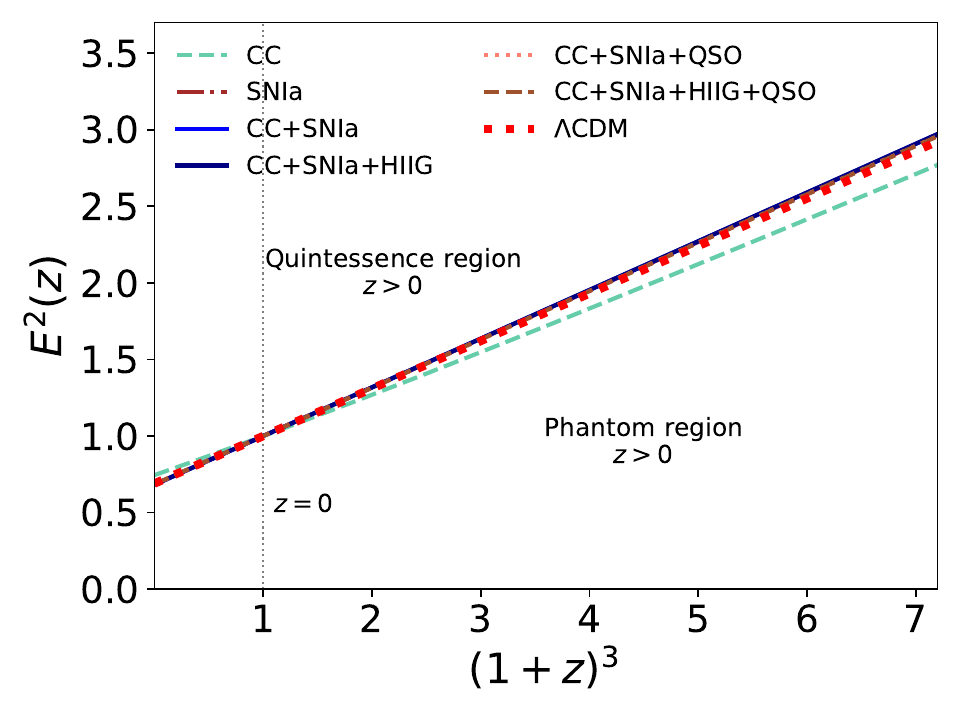}
    \caption{$E^2(z)$ vs $(1+z)^3$ for the QCD model for different data combinations. The standard $\Lambda$CDM  is included as red dashed lines.}
    \label{fig:Ez2}
\end{figure}

Figure~\ref{fig:Ez2} displays $E^{2}(z)$ as a function of $(1+z)^{3}$, allowing geometric comparison between the model and the $\Lambda$CDM, highlighting regions where DE behaves as a cosmological constant, quintessence, or a phantom field. Notably, subtle differences arise: the Universe appears to emerge from a phantom field (under the CC-only constraint) or quintessence (under other constraints) at $z > 0$, converging to a cosmological constant at $z = 0$. In the future ($z < 0$), the model suggests a tendency towards quintessence-like behavior.

The parameter $d$ controls the scaling behavior of the QCD-induced energy density term $\rho_H(z) \propto H^d$. Our analysis reveals that the best-fit values for $d$ agree with values near zero, effectively reproducing the standard expansion history of the $\Lambda$CDM while introducing subtle but observationally significant deviations. 
A positive value of $d$ indicates that this term dilutes with cosmic time, corresponding to a transient or emergent dark energy component that played a more dominant role in the past. In contrast, a slightly negative $d$ produces a slowly growing contribution that could eventually dominate the future expansion, potentially driving a phase of superacceleration. 
Remarkably, our constrained best-fit value, $d = -0.004^{+0.027}_{-0.057}$, remains sufficiently small to prevent late-Universe instabilities or divergences. This result strongly suggests that the proposed QCD coupling in our model acts as a stable dynamical modification to the vacuum energy, with observational consequences that remain consistent with current cosmological data.

Finally, as a statistical comparison between the QCD-modified model and the $\Lambda$CDM, we use Akaike's Information Criterion (AIC) \cite{AIC:1974, Sugiura:1978}, defined as:
\begin{equation}
    \text{AIC} \equiv \chi^2 + 2k,
\end{equation}
where $\chi^2$ is the chi-square of the best fit value and $k$ is the number of free parameters (fitted). The preferred model is the one with the lowest AIC value. The interpretation of the AIC difference ($\Delta$AIC) is as follows:
\begin{itemize}
    \item If $\Delta$AIC $< 4$, both models are equally supported by the data.
    \item If $4 < \Delta$AIC $< 10$, the data still support the given model but less than the preferred~one.
    \item If $\Delta$AIC $> 10$, the observations do not support the given model.
\end{itemize}
Furthermore, we compute the Bayesian Information Criterion (BIC) \cite{schwarz1978}, defined as:
\begin{equation}
    \text{BIC} \equiv \chi^2 + k \log(N),
\end{equation}
where $N$ is the sample size. The BIC imposes a stronger penalty on model complexity than the AIC. Similarly to AIC, the best model corresponds to the lowest BIC value. The interpretation of the difference in BIC ($\Delta$BIC) is as follows:
\begin{itemize}
    \item If $\Delta$BIC $< 2$, there is no significant evidence against the model.
    \item If $2 < \Delta$BIC $< 6$, there is modest evidence against the candidate model.
    \item If $6 < \Delta$BIC $< 10$, the evidence against the model is strong.
    \item If $\Delta$BIC $> 10$, the evidence against the model is very strong.
\end{itemize}

Table~\ref{tab:AICBIC} shows the values of $\Delta$AIC=AIC$^{\rm QCD}$-AIC$^{\Lambda\rm CDM}$ and $\Delta$BIC=BIC$^{\rm QCD}$-BIC$^{\Lambda\rm CDM}$. In this notation,  negative values of these differences mean that the lowest value of AIC (BIC) corresponds to the QCD-modified cosmology. According to the AIC, both models are equally supported by the combined datasets. However, when applying a stronger penalization for model complexity, as implemented by the BIC, only the SNIa dataset provides evidence disfavoring the QCD-modified model, while for the remaining datasets the conclusions remain consistent with those obtained from the AIC analysis.

According to the AIC and BIC, both models are equally supported by multiple datasets. 

\begin{table}[h]
        \caption{Statistical comparison between the QCD-modified model and $\Lambda$CDM using Akaike and Bayesian Information Criteria. $\Delta$AIC = AIC$^{\rm QCD}$ $-$ AIC$^{\Lambda\rm CDM}$ and $\Delta$BIC = BIC$^{\rm QCD}$ $-$ BIC$^{\Lambda\rm CDM}$ represent the difference between the QCD cosmology and the $\Lambda$CDM values. Negative values of $\Delta$ represent a preference for the QCD-modified cosmology.}    \label{tab:AICBIC}
\begin{tabularx}{\linewidth}{lcccccc}
\toprule
     \textbf{Data}              & \textbf{AIC (QCD)} & \textbf{AIC (\boldmath{$\Lambda$}CDM)} & \textbf{\boldmath{$\Delta$}AIC} & \textbf{BIC (QCD)} & \textbf{BIC (\boldmath{$\Lambda$}CDM)} & \textbf{\boldmath{$\Delta$}BIC}  \\
\midrule
     CC                &   22.52 &	25.15	& $-$2.63	& 26.92	    & 28.08	    & $-$1.16 \\
     SNIa              & 2017.65 &	2015.44	&  2.21	& 2033.97	& 2026.32   &  7.65 \\
     CC+SNIa           & 2032.57 &	2036.37	& $-$3.8	& 2048.94	& 2047.28   &  1.66 \\
     CC+SNIa+HIIG      & 2473.77 &	2477.63	& $-$3.86	& 2490.44	& 2488.74	&  1.70 \\
     CC+SNIa+QSO       & 5203.86 &	5207.69	& $-$3.83	& 5220.43	& 5218.74	&  1.69 \\
     CC+SNIa+\\HIIG+QSO  & 5643.75 &	5647.6	& $-$3.85	& 5660.60	& 5658.83	&  1.77 \\
\bottomrule
\end{tabularx}
\end{table}

\subsection{QCD Results}

We now examine the impact of  cosmological coupling on the QCD thermodynamics predicted by the modified PNJL model. Before presenting the results, it is important to clarify the methodology, since the QCD sector operates at a vastly different energy scale from cosmology and all parameters must be fixed consistently.

\subsubsection{Methodology  and Parameter Fixing}

All PNJL parameters entering the standard Polyakov-loop potential $\mathcal{U}(\Phi,\Phi^{*},T)$—namely $a_{0}$, $a_{1}$, $a_{2}$, $a_{3}$, $b_{3}$, $b_{4}$ and $T_{0}$—are kept fixed to their usual values obtained from lattice QCD calibrations. No cosmological observation modifies these quantities. 

The only modification to the thermodynamic potential is the additional term in \mbox{Equation~\eqref{eq:deltaU}}, introduced in Equation~\eqref{eq:Uprime}. Since the relevant QCD physics occurs at temperatures
of order $T \sim 100$--$300\ \mathrm{MeV}$, whereas the Hubble parameter at the present epoch
is extremely small $H_0 \sim 10^{-33}\,\mathrm{eV}$, the cosmic expansion may be treated as
effectively static from the perspective of QCD. As explained in Section~\ref{sec:Adiabatic}, this justifies
an adiabatic approximation in which $H$ acts as an external input.

With all PNJL parameters fixed and only $d$ varied, the gap equations are solved to obtain the temperature dependence of the relevant order parameters—the dynamical quark mass, the chiral condensate,  and the Polyakov loop—and their susceptibilities. These solutions allow us to explore the thermodynamic consequences of the cosmological coupling, including its impact on the thermal evolution of the chiral and Polyakov-loop sectors, the phase transitions, and the CEP, in direct comparison with the standard PNJL model and its finite-volume MRE extensions.

\subsubsection{Thermal Evolution and Phase Transitions}

\begin{figure}[h]
    \includegraphics[width=0.6\textwidth]{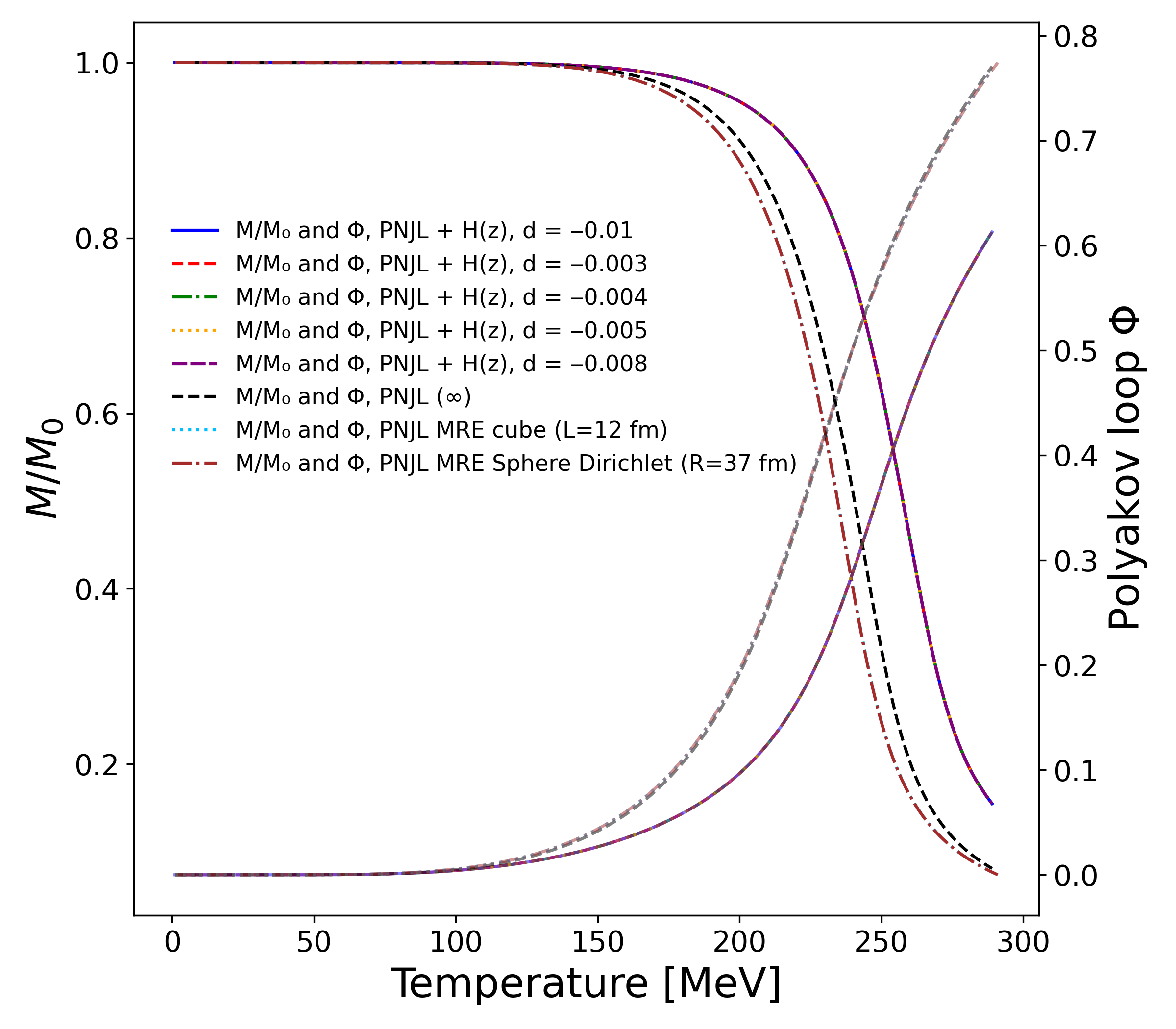}
    \caption{Chiral condensate and Polyakov loop as a function of temperature for the modified PNJL~model.}
    \label{fig:M and loop vs T}
\end{figure}

Figure \ref{fig:M and loop vs T} displays the thermal behavior of the normalized chiral condensate  ($M/M_0$) and the Polyakov loop ($\Phi$), representing the thermal evolution in different scenarios, the standard PNJL model, the finite volume extensions using the Multiple Reflection Expansion (MRE) for both cubic and spherical geometries, and our modified PNJL model with cosmological coupling. For the finite-volume corrections, we adopt the MRE approach, which modifies the density of states by including surface and curvature contributions. This method has been successfully used in QCD effective models to account for confinement in finite geometries, e.g., in heavy-ion or astrophysical contexts \cite{Saucedo2023, Kiriyama2006}. In our study, the MRE provides a reference scenario to compare with the effects induced by the cosmological~\mbox{coupling}.

As seen in Figure \ref{fig:M and loop vs T}, the transition associated with both chiral restoration and deconfinement occurs in a smooth crossover manner, consistent with QCD at zero chemical potential. The thermal transition associated with chiral symmetry restoration occurs first for finite-volume models, in particular in the MRE Dirichlet configuration for a sphere, as compared to the standard PNJL model in the infinite-volume limit. This behavior agrees with the expected strengthening of the confinement effects in reduced volumes. On the other hand, PNJL models incorporating the $H(z)$ cosmological coupling show slightly delayed transitions, indicating that the expansion of the Universe, modeled by the $H^d$ term, has a smoothing effect on the phase change. Furthermore, variations in the exponent $d$ produce only minimal differences in critical behavior, suggesting that the model is not sensitive to this parameter at $\mu = 0$.

The Polyakov loop, which acts as an order parameter for deconfinement, shows the expected behavior in all models; at low temperatures the parameter is practically zero, and with increases in temperature the parameter starts to grow until approaching unity. Among the different scenarios, the MRE finite-volume models show an earlier rise in $\Phi$, indicating a lower deconfinement temperature, while the standard PNJL model shows a more delayed transition. The models with $H(z)$ cosmological coupling show an even later onset of deconfinement, suggesting that the expansion of the Universe weakens the onset of color deconfinement. Moreover, the curves for different values of the exponent d overlap almost completely, reflecting that the Polyakov loop also depends weakly on this parameter under the conditions considered.

The interplay between quark deconfinement and chiral symmetry restoration remains a central question in QCD thermodynamics. To explore this connection, we also consider the temperature at which the normalized chiral condensate $M/M_0$ intersects the Polyakov loop $\Phi$. This crossing provides a qualitative indicator of the relative behavior of both order parameters, rather than a precise definition of the transition temperature. As shown in Table \ref{tab:Intersection_M_Phi_full}, the intersection occurs within a narrow temperature range between 240 and 260 MeV in all models. These values are model-dependent and reflect the specific PNJL parametrization employed, rather than the physical QCD crossover temperature. The near coincidence illustrates the close correlation between the chiral and deconfinement sectors under the conditions studied. The intersection values lie between 0.45 and 0.50, indicating a partial restoration of both order parameters, consistent with a smooth crossover at zero chemical potential.

\begin{table}[h]
     \caption{Characteristic temperatures obtained within the PNJL framework for different model configurations. $T_{\text{int}}$ denotes the temperature at which the normalized chiral condensate $M/M_0$ and the Polyakov loop $\Phi$ intersect. The column $\langle \bar{q} q \rangle_{\text{int}} = \Phi_{\text{int}}$ reports the common value at that intersection point. These values are model-dependent and are included for comparative purposes.}
    \label{tab:Intersection_M_Phi_full}
    \begin{tabularx}{\linewidth}{lcc}
        \toprule
        \textbf{Model} & \boldmath{$T_{\textbf{\text{int}}}$} \textbf{[MeV]} & \boldmath{$\langle \bar{q} q \rangle_{\textbf{\text{int}}} = \Phi_{\textbf{\text{int}}}$}         \\
        \midrule
        PNJL + $H(z)$ ($d = -0.01$) & 260.17 & 0.455 \\
        PNJL + $H(z)$ ($d = -0.003$) & 260.27 & 0.453 \\
        PNJL + $H(z)$ ($d = -0.004$) & 260.17 & 0.455 \\
        PNJL + $H(z)$ ($d = -0.005$) & 260.20 & 0.454 \\
        PNJL + $H(z)$ ($d = -0.008$) & 260.37 & 0.451 \\
        PNJL MRE cube (L = 12 fm) & 236.24 & 0.470 \\
        PNJL MRE Dirichlet (R = 37 fm) & 236.13 & 0.471 \\
        PNJL ($\infty$) & 240.27 & 0.504 \\
        \bottomrule
    \end{tabularx}
\end{table}

\begin{figure}[h]
    \includegraphics[width=0.6\textwidth]{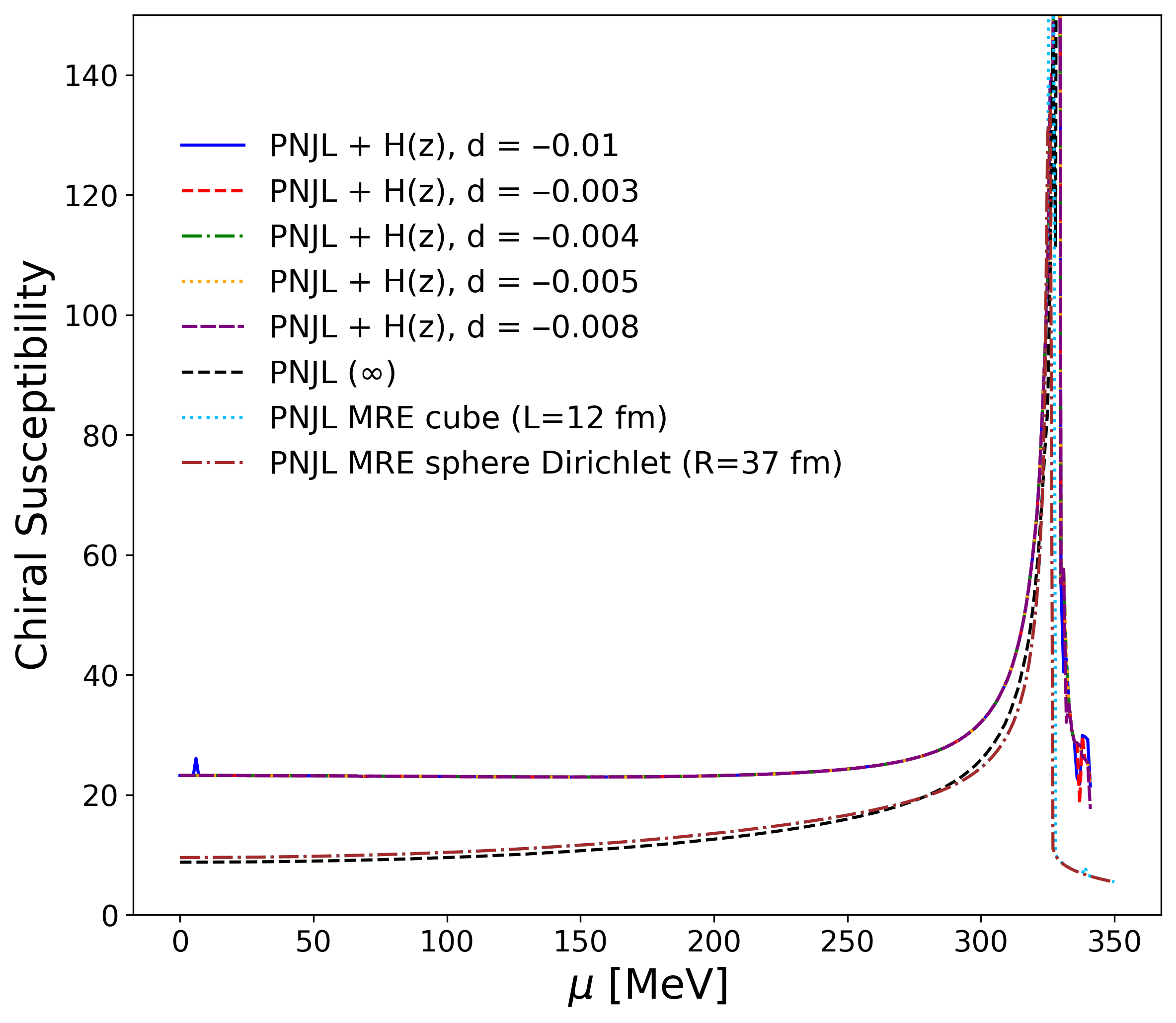}
    \caption{Maximal chiral susceptibility as a function of the chemical potential for the PNJL model in different configurations: finite volume with MRE (cube $L=12$ fm, sphere $R=37$ fm), standard PNJL, and the PNJL model with $H(z)$ coupling.}
    \label{fig:Maximal Chiral Susceptibility vs mu}
\end{figure}

{ Figure \ref{fig:Maximal Chiral Susceptibility vs mu} displays the maximal value of the chiral susceptibility as a function of the chemical potential for different configurations of the PNJL model. This observation provides a sensitive probe of the thermal behavior of the system and is commonly used to explore the phase structure of effective QCD models. The interpretation of these results and their implications for the critical end point are discussed in the following subsection.}

\subsubsection{Critical End Point (CEP)}

To locate the critical point, we follow a geometric criterion inspired by Ref.~\cite{Saucedo2023}, designed to identify the onset of critical behavior in the chiral susceptibility. In effective QCD models, the susceptibility does not exhibit a strict divergence due to finite numerical resolution, but rather develops increasingly pronounced peaks as the system approaches the CEP.
To characterize this behavior, we compute the chiral susceptibility over the $(T,\mu)$ plane and evaluate the local slope between consecutive points along lines of constant chemical potential. When the angular inclination of the susceptibility profile exceeds $89^\circ$, corresponding to a near-vertical growth in the susceptibility, we take this as a numerical indicator of critical-like behavior. The associated $(T,\mu)$ coordinates are then used to identify the CEP region. The angular criterion employed here should therefore be understood as a practical numerical diagnostic of the rapid growth of the chiral susceptibility, rather than as a physical definition of the critical point. Accordingly, the extracted CEP location should be interpreted qualitatively, and with the understanding that its precise value may depend on the specific numerical prescription and resolution adopted.

\begin{figure}[h]
    \includegraphics[width=0.6\textwidth]{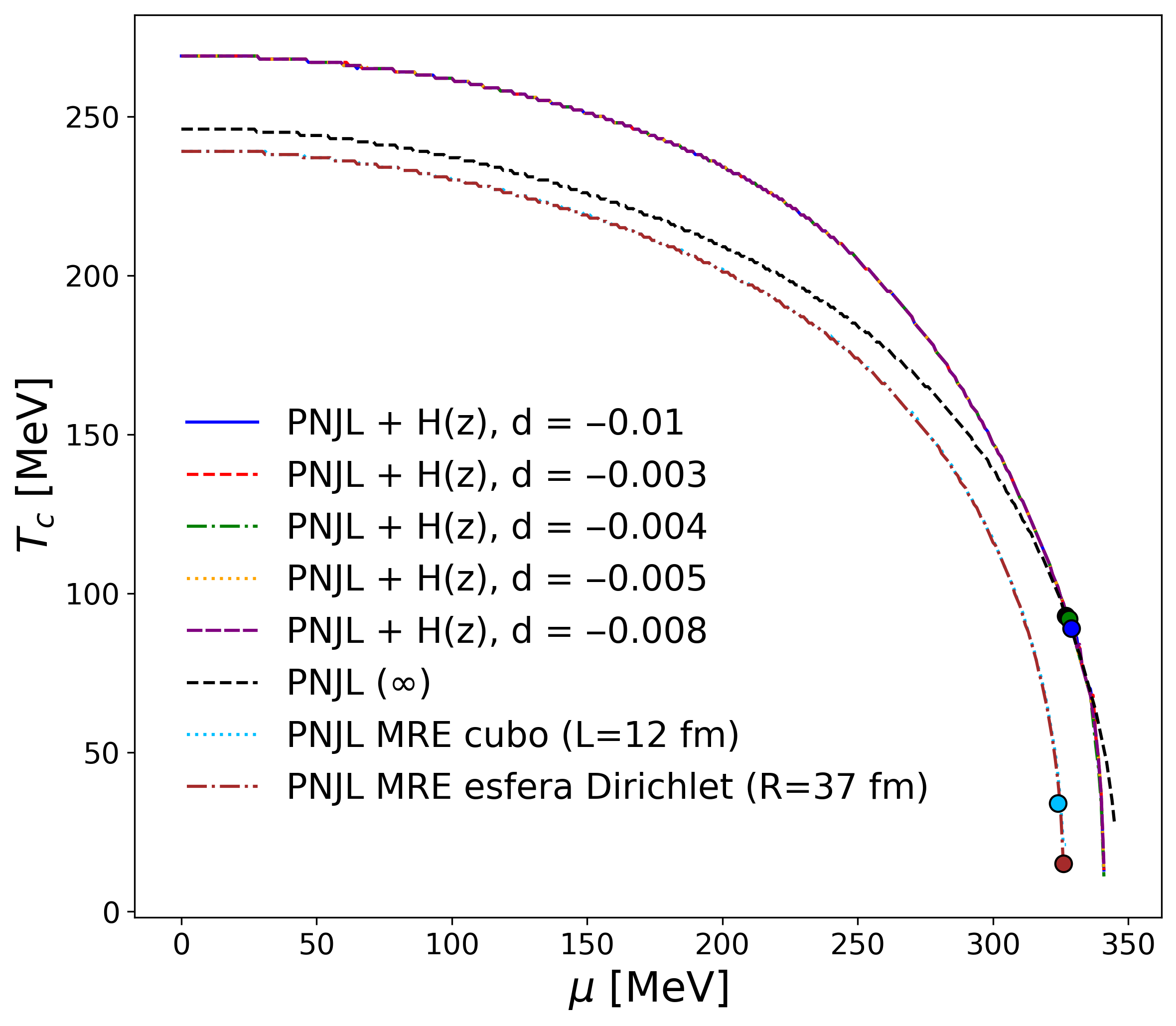}
  \caption{Phase diagram in the \( T_c \)--\( \mu \) plane for different QCD effective models. The solid, dashed, and dash--dotted colored curves represent the confinement--deconfinement transition lines for the PNJL model with cosmological correction \( H(z) \) (for various values of \( d \)). The black curve corresponds to the standard PNJL model, while the cyan and brown curves represent the PNJL model with finite volume corrections using the MRE approximation for a cube (\( L=12~\mathrm{fm} \)) and a sphere with Dirichlet boundary conditions (\( R=37~\mathrm{fm} \)), respectively. Critical end points (CEPs) for each configuration are indicated by colored markers.}
\label{fig:phase_diagram}
\end{figure}

Then, taking the maximum value of the chiral susceptibility for each variation in the parameters of \(T\) and \(\mu\), the phase diagram is constructed (Figure \ref{fig:phase_diagram}). Table \ref{tab:CEP_comparison} summarizes the critical points for each modification, which should be interpreted as qualitative estimates obtained from the numerical diagnostic described above. As can be seen, the curve corresponding to the modified models with coupling \(H(z)\) is shifted toward higher temperatures and chemical potentials, indicating that more extreme conditions are required for symmetry restoration to occur. Despite the overall shift of the transition line in the plane, the location of the CEP for models that include the parameter $H(z)$ appears to be surprisingly close to the value of the standard PNJL model. This suggests that, although the cosmological coupling acts as a stabilizing force that delays the restoration of chiral symmetry and deconfinement, it does not substantially alter the local curvature or the location of the critical point.

These results are consistent with our theoretical motivation behind the structure associated with \(H(z)\). Recall that the function \(f(\Phi)\) was constructed to enhance the cosmological influence in the confined phase (\(\Phi \approx 0\)) and suppress it in the deconfined phase (\(\Phi \approx 1\)). As a result, we observe a significant shift in the phase boundary at low temperatures and chemical potentials, precisely where the vacuum structure of QCD predominates and the influence of cosmic expansion is expected to be strongest.

However, near the CEP, where the system approaches the deconfined regime, the coupling to \(H(z)\) becomes negligible. This provides a natural interpretation for why the critical endpoint remains close to that obtained in the standard PNJL model. This behavior suggests that the parameter associated with the cosmological expansion $H(z)$ affects the dynamics of QCD, but has little impact in the high-energy perturbative regime.

\begin{table}[h]
    \caption{Coordinates of the critical end point (CEP) for different variants of the PNJL model.}
    \label{tab:CEP_comparison}
    \begin{tabularx}{\linewidth}{lccc}
        \toprule
        \textbf{Model} & \boldmath{$T_{\textbf{\text{CEP}}}$} \textbf{[MeV]} & \boldmath{$\mu_{\textbf{\text{CEP}}}$} \textbf{[MeV]} \\
        \midrule
         PNJL + $H(z)$ ($d = -0.01$) & 89 & 329 \\
        PNJL + $H(z)$ ($d = -0.003$) & 92 & 328 \\
        PNJL + $H(z)$ ($d = -0.004$) & 92 & 328 \\
        PNJL + $H(z)$ ($d = -0.005$) & 89 & 329 \\
        PNJL + $H(z)$ ($d = -0.008$) & 89 & 329 \\
        PNJL MRE cube (L = 12 fm) & 34 & 324 \\
        PNJL MRE Dirichlet (R = 37 fm) & 15 & 326 \\
        PNJL ($\infty$) & 93 & 327 \\
        \bottomrule
\end{tabularx}
\end{table}

Overall, the QCD results confirm that the cosmological coupling behaves as expected: it modifies the vacuum structure in the confined phase while leaving the high-temperature regime and the critical end point largely unchanged.

\section{Conclusions and Discussions} \label{SD}

In this work, we have explored a phenomenological extension of the PNJL model in which the Polyakov-loop potential is coupled to the cosmic expansion rate through a term of the form $(H/H_{0})^{d}f(\Phi,\Phi^{*})$. This modification aims to capture the possibility that non-perturbative QCD vacuum effects in the confined phase may carry a residual imprint of the large-scale dynamics of the Universe. The construction preserves the thermodynamic consistency of the PNJL framework, requires only a single new exponent $d$, and is built to vanish naturally in the deconfined regime where $f(\Phi,\Phi^{*}) \to 0$. The introduction of the coupling term  $f(\Phi,\Phi^{*})$  suppresses contributions in the deconfined regime and allows the model to effectively mimic the dark energy at late times without requiring it to be fundamental or constant. This may provide an alternative perspective on long-standing theoretical challenges, such as the cosmological constant problem~\cite{Zeldovich,Weinberg,Carroll:2000}, and aligns with recent observational tensions pointing toward deviations from a pure $\Lambda$CDM model, including those reported by ~\cite{Zhao:2017cud, DiValentino:2020zio, Hernandez-Almada:2020uyr} and other studies~\cite{DESI:2025zgx}.

We have also shown how this coupling leads to an effective dynamical dark-energy contribution $\zeta E^{d}(z)$ in the Friedmann equation, where the constant $\zeta = CH_0^{d-2}$ arises directly from the modified Polyakov potential. This establishes a clear and explicit connection between the QCD sector and the cosmological expansion. The resulting cosmology reduces to $\Lambda$CDM
when $d = 0$, while allowing for a controlled deviation determined by the exponent~$d$.

Using a combination of low-redshift datasets (cosmic chronometers, Type Ia supernovae, HII galaxies and quasars), we find that the observational constraints favor values of $d$ that are statistically consistent with zero but still allow for small departures from the $\Lambda$CDM at the $1\sigma$ level. The reconstructed expansion history shows that the model behaves very similarly to a cosmological constant at present, while permitting mild quintessence- or phantom-like trends at intermediate redshifts. Cosmographic diagnostics confirm that these deviations remain small and compatible with current uncertainties.

The QCD sector exhibits behavior consistent with the expectations of effective models. The cosmological coupling slightly delays both chiral restoration and deconfinement transitions by stabilizing the confined vacuum, yet the overall structure of the crossover remains intact. Importantly, within the qualitative resolution of the present numerical analysis, the location of the critical end point remains close to that obtained in the standard PNJL model. This is a direct consequence of the suppression factor $f(\Phi,\Phi^{*})$, which ensures that the coupling has negligible impact in the deconfined regime where the CEP resides.

From an observational point of view, the model performs comparably to the $\Lambda$CDM in fitting low-redshift observables (SNIa, CC), while introducing a physically motivated dynamical component that vanishes at early times. This behavior avoids potential tensions with Big Bang nucleosynthesis and the CMB, provided the function $f(\Phi)$ effectively suppresses early-time contributions. Additionally, an interesting behavior is observed in \mbox{Figure \ref{fig:Ez2}}, where a past Phantom behavior is subtly observed in CC constraints, having an interesting correlation with the recent results from the Dark Energy Spectroscopic Survey (DESI) collaboration \cite{DESI:2025zgx}. Meanwhile, for the other constraints it is observed that quintessence behavior and a tendency to a cosmological constant at $z=0$ are expected; finally, a future ($z<0$) dominance of quintessence ( $z<0$) is shown according to Figure \ref{fig:Ez2}. We also observe from the AIC and BIC results that the $\Lambda$CDM and QCD cosmology are equally supported for most of the datasets considered, with the exception of the single-SNIa sample, for which the BIC indicates strong statistical evidence against the QCD cosmological model. Furthermore, from the 2D contours of the fitted parameters (see Figure \ref{fig:contours} and Table \ref{tab:bf_model}), we can observe that the constraints of the extra parameter $d$ are in agreement with the case $d=0$, in which the $\Lambda$CDM is recovered, suggesting that there is no  positive evidence for this extended model for the present low-redshift data.

From a phenomenological standpoint, this framework offers a tractable path to bridge the non-perturbative QCD vacuum with cosmological dynamics. Future developments could investigate whether such coupling terms arise naturally from effective QCD theories in curved spacetime or from holographic QCD constructions adapted to expanding backgrounds. Additionally, lattice QCD simulations incorporating curvature or time-dependent metrics may provide insight into the behavior of confinement under cosmological conditions. Upcoming high-precision cosmological surveys, such as Euclid \cite{Amendola:2016saw} and LSST \cite{Blum:2022dxi}, may also test the predictions of the model, particularly with regard to the evolution of the Hubble parameter and deviations from the $\Lambda$CDM at intermediate redshifts.

From a theoretical point of view, the coupling term is phenomenologically motivated. Although it preserves the dimensional structure of the Polyakov potential and aligns with the qualitative behavior expected in confined QCD phases, its origin from first-principles QCD or quantum field theory in curved spacetime remains to be developed. Possible directions include deriving the correction from the effective action of QCD in curved FLRW backgrounds or from holographic QCD models under cosmological conditions

Overall, the present work provides a consistent and observationally viable link between QCD confinement physics and the late-time acceleration of the Universe, offering a minimal and testable alternative to a fundamental cosmological constant.

\vspace{6pt} 






\acknowledgments{We thank the anonymous referees for their thoughtful remarks and suggestions. We thank the support of  J.R. Morones Ibarra for the improvement of the paper. A.H.A. is grateful for  the support from Luis Aguilar, Alejandro de Le\'on, Carlos Flores, and Jair Garc\'ia of the Laboratorio  Nacional de Visualizaci\'on Cient\'ifica Avanzada. M.A.G.-A. acknowledges support from c\'atedra Marcos Moshinsky, SECIHTI for the support with the National Research System (SNII) grant and the project 0056 from Universidad Iberoamericana: Nuestro Universo en Aceleraci\'on, energ\'ia oscura o modificaciones a la relatividad general. The numerical analysis was also carried out by the  Numerical Integration for Cosmological Theory and Experiments in High-energy Astrophysics (Nicte Ha) cluster at IBERO University, acquired through c\'atedra MM support. A.H.A and M.A.G.-A acknowledge partial support from project ANID Vinculaci\'on Internacional FOVI220144. H.M-H. acknowledges support from SECIHTI CBF2023-2024-1630.}


\end{document}